\documentclass[aps,prd,twocolumn,groupedaddress,amssymb]{revtex4}

\def\plb#1{Phys.~Lett.~{\bf B#1}}

\def\prl#1{Phys.~Rev.~Lett.~{\bf #1}}
\def\prd#1{Phys.~Rev.~{\bf D#1}}

\begin{document}

\preprint{UCD-04-21}

\title{Neutrino Masses and Mixing, Quark-lepton Symmetry  and Strong Right-handed Neutrino 
Hierarchy}

\author{Radovan Derm\' \i\v sek}
\email[]{dermisek@physics.ucdavis.edu}

\affiliation{Davis Institute for High Energy Physics,
University of California, Davis, CA 95616, U.S.A.}

\date{October 26, 2004}

\begin{abstract}

Assuming the same form of all mass matrices as motivated by quark-lepton symmetry, 
we discuss conditions under which bi-large mixing in the
lepton sector can be obtained with a minimal amount of fine tuning requirements for possible models.
We assume hierarchical mass matrices, dominated by the 3-3 element, with off-diagonal elements much smaller 
than the larger neighboring diagonal element. 
Characteristic features of this scenario are
strong hierarchy in masses of right-handed neutrinos, and 
comparable contributions of both
lighter right-handed neutrinos to the resulting left-handed neutrino Majorana mass matrix.
Due to obvious quark lepton symmetry, this approach can be  embedded into grand unified
theories.
The mass of the
lightest neutrino does not depend on details of a model in the leading order. 
The right-handed neutrino scale
can be identified with the GUT scale in which case
the mass of the lightest neutrino
is given as $(m_{top}^2/M_{GUT}) \, | \, U_{\tau 1} \, |^2$.

\end{abstract}


\maketitle






\section{Introduction}

A global analysis of neutrino oscillation data~\cite{Gonzalez-Garcia:2003, Maltoni:2003}
gives the best fit to the neutrino mass-squared differences:
$\Delta m^2_{sol}  \equiv m_{\nu_2}^2 - m_{\nu_1}^2  \simeq  6.9 \times 10^{-5} {\rm eV}^2$ 
and
$\Delta m^2_{atm}  \equiv m_{\nu_3}^2 - m_{\nu_1}^2  \simeq  2.6 \times 10^{-3} {\rm eV}^2$,
which, in the case $m_{\nu_1} \ll m_{\nu_2}, \, m_{\nu_3}$, can be interpreted as:
\begin{eqnarray}
m_{\nu_2} & \simeq & \sqrt{\Delta m^2_{sol}} \; \simeq  8.3 \times 10^{-3} {\rm eV}, \label{eq:mnu2} \\
m_{\nu_3} & \simeq & \sqrt{\Delta m^2_{atm}} \; \simeq  5.1 \times 10^{-2} {\rm eV}. \label{eq:mnu3}
\end{eqnarray}
%
%
The $3 \sigma$ ranges for 1-2 and 2-3 mixing angles are:
\begin{eqnarray}
0.23 \leq & \sin^2 \theta_{sol} & \leq 0.39 , \label{eq:sol_3s} \\
0.31 \leq & \sin^2 \theta_{atm} & \leq 0.72 , \label{eq:atm_3s}
\end{eqnarray}
and the $3 \sigma$ upper bound on the 1-3 mixing angle is:
\begin{equation}
\sin^2 \theta_{13} \; \leq \; 0.054 \, \label{eq:13_3s} .
\end{equation}


In grand unified theories [GUTs] both quarks and leptons
originate from the same multiplets of unified gauge symmetry. For example, one generation of standard model 
quarks and leptons together with the right-handed neutrino naturally fits into $16$ dimensional representation of 
SO(10). 
Therefore, one would like to relate the results above 
to what we observe in the quark sector and understand them as a consequence of 
assumptions, the same for both quarks and leptons, 
we make about the structure of mass matrices.
However we know that
mixing angles in the quark sector are small. 
The Cabibbo-Kobayashi-Maskawa matrix $V_{CKM}$ is close to the identity matrix with off-diagonal elements dominated by 
small Cabibbo angle (1-2 mixing). On the other hand, as we see from the results above, mixing in the lepton sector is 
large. The 2-3 mixing is 
close 
to maximal, 1-2 mixing is large, somewhat smaller than maximal, and 1-3 mixing is small, close to zero.
There is no obvious quark-lepton
symmetry in this pattern which makes the unified understanding of mass matrices quite challenging.

It has been realized that the lepton mixing can be enhanced by the see-saw mechanism even if the Dirac Yukawa 
couplings of leptons have a similar structure to that in the quark sector~\cite{smirnov,altarelli_rev}.

In a democratic approach (where all elements of mass matrices are equal in the leading order), it has been shown that 
a bi-large mixing can be obtained while assuming the same form of all 
Yukawa matrices and the right-handed neutrino Majorana mass matrix~\cite{branco_efd, dem1}. 
Even the form of perturbations can be the same for all mass matrices. This makes the approach 
manifestly quark-lepton symmetric and so it can be embedded into GUT models. 
The lepton mixing matrix is 
predominantly given by the matrix diagonalizing the charged lepton Yukawa matrix. Furthermore, there exist a well 
defined 
framework (without exactly specifying perturbations) in which the left-handed neutrino mass matrix contributes the 
minimal amount of mixing to the lepton mixing 
matrix. In this case the lepton mixing matrix is given in terms of two parameters (neglecting phases) and the value of 
one mixing angle, $\sin \theta_{13}$, can be predicted~\cite{dem1}.
If embedded into grand unified theories 
the third generation Yukawa
coupling unification is a generic feature (without necessity of distinguishing the third generation from the other two by 
family 
symmetries or in any other way) while masses of the first two generations of charged fermions depend on
small perturbations. In the neutrino sector, the heavier two neutrinos are model dependent, while the mass of the
lightest neutrino in this approach does not depend on perturbations in the leading order.
Finally, the right-handed neutrino mass scale can
be identified with the GUT scale
in which case
the mass of the lightest neutrino
is given as $(m_{top}^2/M_{GUT})  \, | \, U_{\tau 1} \, |^2 $~\cite{dem1}.
This framework has everything one could wish for: obvious quark-lepton symmetry, 3rd generation Yukawa coupling 
unification, bi-large lepton mixing
with a prediction for $\sin \theta_{13}$ in the minimal case, 
and more importantly, 
no need for an intermediate right-handed neutrino scale and with that associated prediction for the mass of the 
lightest neutrino. There is one problem however: it is not straightforward to build a concrete model with the usual use 
of 
family symmetries (for discussion see~\cite{dem1}).

A similar approach, also  assuming the same form of all mass matrices in the leading order
was recently discussed in
Ref.~\cite{dorsner_smirnov}. Instead of democratic mass matrices, the starting point is a singular matrix of the form
$(\lambda^2, \lambda, 1 )^T . \, (\lambda^2, \lambda, 1 )$.

Motivated by these finding we want to identify the corresponding picture in the hierarchical approach.
Afterall, a democratic mass matrix (and a matrix of the type $(\lambda^2, \lambda, 1 )^T . \, (\lambda^2, \lambda, 1 
)$ as well)
is equivalent to a matrix with the 3-3 element only. This is the usual starting point 
of hierarchical models. 

In this letter
we assume hierarchical mass matrices, dominated by the 3-3 element, with off-diagonal elements much smaller
than the larger neighboring diagonal element:
\begin{equation}
Y_f  \sim \left(\begin{array}{ccc} 0 & \delta_f & \delta_f \\
                         \delta_f & \epsilon_f & \epsilon_f \\
                         \delta_f & \epsilon_f & 1
                   \end{array} \right) \,  \lambda_f \, , \quad f = u, d, e, \nu,
\label{eq:texture}
\end{equation}
where $0 \ll \delta_f \ll \epsilon_f \ll 1$ represent only orders of magnitude of different elements of Yukawa matrices.
Clebsch-Gordan coefficients (or order one couplings) necessary to explain quark-lepton mass relations 
of the 
first two generations are understood. Yukawa matrices of this type naturally explain hierarchy in masses of 
charged fermions and mixing in the quark sector. 
Rather than starting with a specific model or texture 
we discuss conditions under which bi-large mixing in the  
lepton sector can be obtained assuming the same generic structure of the neutrino Yukawa matrix.
We find that the characteristic feature of this scenario is a strong hierarchy in masses of right-handed neutrinos and 
comparable contributions of both
lighter right-handed neutrinos to the resulting left-handed neutrino mass matrix.
Furthermore, the heaviest right-handed neutrino have to contribute negligibly
which 
leads to a prediction for the mass of the lightest neutrino.
The right-handed neutrino scale
can be identified with the GUT scale in which case
(assuming third generation Yukawa coupling unification) the mass of the lightest neutrino
is given as $(m_{top}^2/M_{GUT}) \, | \, U_{\tau 1} \, |^2$. It does not depend on details of a model ($\epsilon$'s and 
$\delta$'s).
Finally, we discuss how suitable right-handed neutrino sector can be naturally obtained in SO(10) models. 

\section{Conditions for bi-large lepton mixing}

We assume that the right-handed neutrino Majorana mass matrix has the same hierarchical structure as Yukawa matrices
(this is not necessarily required by quark-lepton symmetry, but it is well motivated, see the discussion later).
For simplicity, we work in the basis where the right-handed neutrino Majorana mass 
matrix is 
diagonal, defined as
\begin{equation}
M_{\nu_R} =      \left(\begin{array}{ccc}  r_1 & 0 & 0 \\
                                     0 & r_2 & 0 \\
                                     0 & 0 & 1 \\
                 \end{array} \right) \, M_0 \, ,
\end{equation}
where $r_1 \ll r_2 \ll 1$, and $M_0$ is the scale at which right-handed neutrino masses are generated. Later 
we take $M_0 = M_{GUT}$. In this basis the neutrino Yukawa matrix (defined with doublets on the left) 
is in general given as
\begin{equation}
Y_\nu  = \left(\begin{array}{ccc} \epsilon_{11} & \epsilon_{12} & \epsilon_{13} \\
                                      \epsilon_{21} & \epsilon_{22} & \epsilon_{23} \\
                                      \epsilon_{31} & \epsilon_{32} & 1
                   \end{array} \right) \,  \lambda_\nu \, ,
\end{equation}
where we assume the same order of perturbations $\epsilon_{ij}$ as in Eq.~(\ref{eq:texture}), namely
$\epsilon_{11} \ll \epsilon_{21} \sim \epsilon_{31} \ll \epsilon_{22} \sim \epsilon_{32} \ll 1$ with 
$\epsilon_{ij} \sim \epsilon_{ji}$.
The inverse of the  right-handed neutrino Majorana mass  
matrix is given as:
\begin{equation}
M_{\nu_R}^{-1} = \frac{1}{r_1 r_2 M_0} \,  
             \left(\begin{array}{ccc} r_2 & 0 & 0 \\
                                      0 & r_1 & 0 \\
                                      0 & 0 & r_1 r_2
                   \end{array} \right).
\end{equation}
When right-handed neutrinos are integrated out we obtain the left-handed neutrino Majorana mass matrix 
given by the see-saw formula~\cite{see-saw}:
\begin{equation}
M_{\nu_L} = - v^2_u \, Y_\nu M_{\nu_R}^{-1} Y_\nu^T .
\label{eq:see-saw}
\end{equation}
In our basis it can be written in three terms, each  corresponding to the contribution of one right-handed 
neutrino:
\begin{equation}
M_{\nu_L} = - \frac{\lambda^2 v^2_u}{r_1 r_2 M_0} \left( 
{\cal M}_1 + {\cal M}_2 + {\cal M}_3
\right) ,
\end{equation}
where
\begin{eqnarray}
{\cal M}_1 & = & r_2  \left(\begin{array}{ccc} 
   \epsilon_{11}^2 & \epsilon_{11} \epsilon_{21} & \epsilon_{11} \epsilon_{31} \\
   \epsilon_{21} \epsilon_{11} & \epsilon_{21}^2 & \epsilon_{21} \epsilon_{31} \\
   \epsilon_{31} \epsilon_{11} & \epsilon_{31} \epsilon_{21} & \epsilon_{31}^2
                   \end{array} \right) 
 =   r_2 \, {\vec e}_1 \, . \, {\vec e}^{\; T}_1  ,  \\
{\cal M}_2 & = & r_1   \left(\begin{array}{ccc}  
   \epsilon_{12}^2 & \epsilon_{12} \epsilon_{22} & \epsilon_{12} \epsilon_{32} \\
   \epsilon_{22} \epsilon_{12} & \epsilon_{22}^2 & \epsilon_{22} \epsilon_{32} \\
   \epsilon_{32} \epsilon_{12} & \epsilon_{32} \epsilon_{22} & \epsilon_{32}^2
                   \end{array} \right)
 =   r_1 \, {\vec e}_2 \, . \, {\vec e}^{\; T}_2  , \\
{\cal M}_3 & = & r_1 r_2   \left(\begin{array}{ccc}  
   \epsilon_{13}^2 & \epsilon_{13} \epsilon_{23} & \epsilon_{13}  \\
   \epsilon_{23} \epsilon_{13} & \epsilon_{23}^2 & \epsilon_{23}  \\
   \epsilon_{13}  & \epsilon_{23}  & 1
                   \end{array} \right)
 =   r_1 r_2 \, {\vec e}_3 \, . \, {\vec e}^{\; T}_3   , 
\end{eqnarray}
and ${\vec e}_i$ is the i-th column of the neutrino Yukawa matrix. Each of these three matrices have two zero 
eigenvalues and a combination of any two of them still has one zero eigenvalue.
The left-handed neutrino mass matrix can be diagonalized by a single unitary matrix,
\begin{equation}
{M}_{\nu_L}^{diag} = U_{\nu_L} M_{\nu_L} U_{\nu_L}^T .
\end{equation}  
Finally, the lepton mixing matrix which appears in the charged current Lagrangian is given as:
\begin{equation}
U = U_e \, U_{\nu_L}^\dag,
\label{eq:U}
\end{equation}
where $U_e$ is the matrix diagonalizing the charged lepton Yukawa matrix, $Y_e^{diag} = U_e Y_e V_e^\dag$. Due to 
the hierarchical nature of the charged lepton Yukawa matrix, $U_e \simeq 1$, and we will neglect it in our discussion.

\subsection{Step 1: The dominant mass matrix}

It is clear that if ${\cal M}_3$ dominates it is not possible to achieve large 2-3 mixing under the 
assumption of 3-3 dominance. The maximal 2-3 mixing corresponds to $\epsilon_{23} = 1$ which goes 
against our motivation. Several models of this type, usually called ``lopsided",  were 
constructed~\cite{lopsided_models}. 
The right-handed neutrino scale $M_0$ 
cannot be identified 
with the GUT scale, since the resulting mass of the heaviest neutrino would be too small, 
$m_{\nu_3} \sim m_{top}^2/M_{GUT} \sim 10^{-3}$ eV. 

Let us assume that ${\cal M}_2$ matrix is the dominant contribution to $M_{\nu_L}$ and let us neglect 
for a moment ${\cal M}_1$ and ${\cal M}_3$. In this case  maximal 2-3 mixing corresponds to 
$\epsilon_{32} = \epsilon_{22}$. This is actually quite good, since $\epsilon_{32}$ and 
$\epsilon_{22}$ are of the same order of magnitude in many models. 
At this
point only the mass of the heaviest neutrino is generated.
The eigenvalues of ${\cal M}_2$ are $\{ 0,0, r_1 \, | \, {\vec e}_2 \, |^2 \}$. 
The eigenvector 
corresponding to $m_{\nu_3}$ is ${\vec e}_2$, and since $U \simeq U_{\nu_L}^\dag$ it will appear (properly normalized) in 
the third 
column of the lepton mixing matrix. 
As a consequence of  degenerate zero eigenvalues the first two columns of the lepton mixing matrix are 
not uniquely specified. In general they can be any orthogonal linear combinations
of two unit vectors orthogonal to ${\vec e}_2$.
Therefore, at this stage the 1-2 mixing angle is not specified, and comparing ${\vec e}_2$ with the 3rd column of $U$, 
parametrized in general as $(s_{13} , s_{23} c_{13}, c_{23} c_{13})^T$, where $s_{13} \equiv \sin \theta_{13}$ and so 
on,  
the 2-3 and 1-3 mixing angles are 
given as:
\begin{equation}
\tan \theta_{23} = \frac{e_{22}}{e_{32}}, \quad \quad {\rm or,} \quad \quad
\sin^2 \theta_{23} = \frac{e_{22}^2}{e_{22}^2 + e_{32}^2}, 
\end{equation} 
and
\begin{equation}
\sin \theta_{13} = \frac{e_{12}}{| \, {\vec e}_2 \, |}.
\label{eq:M2_13}
\end{equation}

Finally, if ${\cal M}_1$ dominates, the discussion is similar to the case of 
${\cal M}_2$ dominance with  ${\vec e}_2$ being replaced by   ${\vec e}_1$. Namely, the eigenvalues of ${\cal M}_1$ are 
$\{ 
0,0, r_2 \, | \, {\vec e}_1 \, |^2 \}$ and the  2-3 and 1-3 mixing angles are
given as:
\begin{equation}
\tan \theta_{23} = \frac{e_{21}}{e_{31}}, \quad \quad {\rm or, } \quad \quad
\sin^2 \theta_{23} = \frac{e_{21}^2}{e_{21}^2 + e_{31}^2},
\label{eq:M1_23}
\end{equation}
and
\begin{equation}
\sin \theta_{13} = \frac{e_{11}}{| \, {\vec e}_1 \, |}.
\label{eq:M1_13}
\end{equation}
If 1-3 mixing angle turns out to be much smaller than the present experimental upper bound,
Eq.~(\ref{eq:13_3s}), this possibility will become strongly 
favored, since $e_{11}$ can be arbitrarily small, and typically, $e_{11} = 0$ in many models.

\subsection{Step 2: Masses of the heavier two neutrinos and the mixing matrix}

The second eigenvalue is lifted and the lepton mixing matrix is specified when the contribution from the 
sub-leading right-handed neutrino is taken into account. To determine which matrix should be next-to-dominant, let us 
look at the eigenvector corresponding to the massless eigenvalue. 
The eigenvector corresponding to the largest eigenvalue (${\vec e}_1$ if ${\cal M}_1$ 
dominates and ${\vec e}_2$ if ${\cal M}_2$ dominates)
which represents the third column 
of the lepton mixing matrix is ${\vec e} \sim (s_{13},c_{13} s_{23}, c_{13} c_{23})^T$. 
Let ${\vec g}$ be ${\vec e}_i$ in the case ${\cal M}_i$ is next-to dominant. 
The normalized eigenvector corresponding to the zero eigenvalue is perpendicular to both ${\vec e}$ and ${\vec g}$ 
and so it can be written as:
\begin{equation}
{\vec v}_0 = \frac{{\vec g} \times {\vec e}}{|\, {\vec g} \times {\vec e} \, |}.
\end{equation}
This vector will appear in the first column of the lepton mixing matrix. 
The first component of this vector (the 1-1 component of the lepton mixing matrix) is given as:
\begin{equation}
( {\vec v}_0 )_1 = \frac{c_{13} \, \left( g_2 c_{23} - g_3 s_{23} \right) }{|\, {\vec g} \, | \, \sin \alpha },
\label{eq:v_0}
\end{equation}
where $\alpha$ is the angle between ${\vec e}$ and ${\vec g}$,
\begin{equation}
\cos \alpha  = \frac{1}{|\, {\vec g} \, |} \left(  g_1 s_{13} + g_2 c_{13} s_{23} + g_3 c_{13} c_{23}  \right).
\label{eq:ca}
\end{equation}
Comparing $( {\vec v}_0 )_1$ with the 1-1 element of $U$, parametrized in general as $c_{12} c_{13}$,
we can read out $\cos \theta_{12}$, and with the use of 
Eq.~(\ref{eq:ca}) we find
\begin{equation}
\sin \theta_{12} = \frac{g_1 c_{13} - \left( g_2 s_{23} + g_3 c_{23} \right) s_{13}}{| \, {\vec g} \, | \, \sin \alpha },
\label{eq:s12}
\end{equation}
or equivalently,
\begin{equation}
\tan \theta_{12} = \frac{g_1 c_{13} - \left( g_2 s_{23} + g_3 c_{23} \right) s_{13}}{g_2 c_{23} - g_3 s_{23} }.
\label{eq:t12}
\end{equation}
This relation was derived in a different way in Ref.~\cite{SRND}.

Since $s_{23} \sim c_{23}$, $c_{13} \sim 1$ and $s_{13} \ll 1 $, 
in order to generate large 1-2 mixing, we need either $g_1 \simeq g_2 - g_3$ or $ g_2  + g_3 \gg g_2 - 
g_3$ assuming non-zero $s_{13}$. This cannot be satisfied for ${\vec g} = {\vec e}_3$.
In this case we need $\epsilon_{23} \simeq 1$ and we would be  back to lopsided textures. 
So ${\cal M}_3$ cannot be  next-to-dominant for the same reason it cannot be dominant.
On the other hand, both ${\cal M}_1$ and ${\cal M}_2$ can easily satisfy above relations. 
For example,
in the case ${\cal M}_2$ is next-to-dominant, ${\vec g} = {\vec e}_2$, and we need
$\epsilon_{12} \simeq \epsilon_{22} - \epsilon_{32}$, which can be satisfied for $\epsilon_{12} \ll \epsilon_{22}$, and
$\epsilon_{22} \sim \epsilon_{32}$
which we already assume anyway.

Let us summarize. If ${\cal M}_1$ dominates, large 2-3 mixing is generated for $\epsilon_{21} \sim \epsilon_{31} 
\sim \delta$, and at this point, the 1-3 mixing is zero or very small as a consequence
of $\epsilon_{11} \sim 0$. Large 1-2 mixing is generated when sub-dominant ${\cal M}_2$ is taken into account and it 
occurs for $\epsilon_{12} \simeq \epsilon_{22} - \epsilon_{32}$. 
Alternatively, if  ${\cal M}_2$ dominates, large 2-3 
mixing is 
generated for 
$\epsilon_{22} \sim \epsilon_{32} 
\sim \epsilon$, and at this point, the 1-3 mixing is small, $\sin \theta_{13} \simeq \epsilon_{12}/\epsilon_{22} \simeq 
\delta/\epsilon$. 
Large 1-2 mixing is generated when sub-dominant ${\cal M}_1$ is taken into account, and it occurs for $\epsilon_{21} 
\simeq \epsilon_{31}$. In this case $\tan \theta_{12} 
\simeq s_{13} (\epsilon_{21} + \epsilon_{31})/(\epsilon_{21} - \epsilon_{31})$.
Both situations are viable.
However, we will see shortly that it is not possible to generate the observed  hierarchy in neutrino masses
under the assumption of single right-handed neutrino dominance while keeping the structure of mass matrices 
strictly hierarchical as given in Eq.~(\ref{eq:texture}).

The non-zero eigenvalues of ${\cal M}_1 + {\cal M}_2$ can be written as:
\begin{equation}
m_\pm = \frac{1}{2} t \left( 1 \pm \sqrt{1 - \frac{4d}{t^2} } \right),
\label{eq:mpm}
\end{equation}
where
\begin{eqnarray}
t & = & r_2 \, | \, {\vec e}_1 \, |^2 \, + \, r_1 \, | \, {\vec e}_2 \, |^2 \, , \\
d & = & r_1 r_2  \, | \, {\vec e}_1 \times  {\vec e}_2 \, |^2  \, = \, r_1 r_2 \, | \, {\vec e}_1 \, |^2 \, | \, {\vec 
e}_2 \, |^2 \, \sin^2 \alpha ,
\end{eqnarray}
and $\alpha$ is the angle between ${\vec e}_1$ and ${\vec e}_2$.
If ${\cal M}_1$ dominates, $r_1 \, | \, {\vec e}_2 \, |^2 \ll r_2 \, | \, {\vec    
e}_1 \, |^2$  and
\begin{equation}
\frac{m_-}{m_+} \simeq \frac{d}{t^2} \simeq \frac{r_1 \, | \, {\vec e}_2 \, |^2 \, \sin^2 \alpha}{r_2 \, | \, {\vec 
e}_1 \, |^2}.
\end{equation}
Numerically, from Eq.~(\ref{eq:mnu2}) and Eq.~(\ref{eq:mnu3}), we find $m_- / m_+ \simeq 0.16$.   
However, examining Eq.~(\ref{eq:s12}), we see that in order to get large 1-2 mixing,  
\begin{equation}
\sin \alpha \sim \frac{\epsilon_{12}}{\epsilon_{22}}, 
\end{equation}
which is naturally of order $\delta / \epsilon$ and not larger than about 0.2. Therefore, the naturally generated 
hierarchy in masses of the heavier two neutrinos is at least of order 100 
(assuming single right-handed neutrino dominance).  

The tension between generating large 1-2 mixing and mild hierarchy in masses of the heavier two neutrinos can be 
relieved when assuming $\epsilon_{12} \simeq \epsilon_{22}$, while keeping $\epsilon_{12} \simeq 
\epsilon_{22} -
\epsilon_{32}$. For a model of this type, see Ref.~\cite{king_ross}. This solution however goes against our 
motivation and is technically the same as the lopsided texture.

In the next section we propose another solution, which keeps the desired form of mass matrices given in 
Eq.~(\ref{eq:texture}). 
Rather we abandon the idea of 
single right-handed neutrino dominance.

\subsection{Comparable contribution of ${\cal M}_1$ and ${\cal M}_2$}

In the previous two sections we showed that in order to obtain large 2-3 and 1-2 mixing angles it  
is necessary that the contribution of ${\cal M}_3$ is neither dominant nor next-to-dominant.
However, the relative contribution  ${\cal M}_1$ and  ${\cal M}_2$ to the resulting left-handed 
neutrino mass matrix was not important at all. No matter which one dominates, the observed pattern of mixing 
angles can be recovered. In the case ${\cal M}_1$ dominates, interpreting the condition for large 1-2 mixing angle 
as $\epsilon_{12} \simeq 
\epsilon_{22} -  \epsilon_{32}$, with $\epsilon_{12} \ll \epsilon_{22}$ rather than $\epsilon_{12} \sim 
\epsilon_{22} \sim \epsilon_{32}$ coincides with the condition for large 2-3 mixing angle if ${\cal M}_2$ matrix 
dominates.
Therefore, as far as 2-3 mixing is concerned the contribution from ${\cal M}_1$ and ${\cal M}_2$ to the left-handed 
neutrino mass matrix can be equal. The masses of the heavier two neutrinos can be then anything, from highly 
hierarchical to degenerate (which happens when $t$ in~Eq.(\ref{eq:mpm}) is zero). 

Due to mild hierarchy in  masses of the heavier two neutrinos, the simplest possibility is 
that ${\cal M}_1$ only
slightly dominates, $r_2 \, | \, {\vec e}_1 \, |^2 \gtrsim  r_1 \, | \, {\vec e}_2 \, |^2 $. The relations for 
mixing 
angles, Eq.~(\ref{eq:M1_23}) and Eq.~(\ref{eq:M1_13}), are now just a rough approximation. The 2-3 mixing will be 
close to maximal, as far as 
$\epsilon_{21} \sim 
\epsilon_{31}$ and $\epsilon_{22} \sim \epsilon_{32}$. The 1-3 mixing will be roughly given by the larger of 
Eq.~(\ref{eq:M1_13}) and Eq.~(\ref{eq:M2_13}).
And finally, the 1-2 mixing can be again obtained by looking at the first component of the eigenvector 
corresponding to 
the zero eigenvalue. Assuming $\epsilon_{11} = 0$ and $\epsilon_{21} = \epsilon_{31}$ we get
\begin{equation}
{\vec v}_0 \, \propto \, {\vec e}_1 \times  {\vec e}_2 \, = \, \left( \, \epsilon_{32} - \epsilon_{22}, \;
\epsilon_{12}, \; - \epsilon_{12} \, \right)^T ,
\end{equation}
which does not depend on relative dominance of ${\cal M}_1$ and ${\cal M}_2$. Comparing the first component of 
${\vec v}_0$ with the 1-1 element of the lepton mixing matrix,  $c_{12} c_{13}$, and taking   
$\cos \theta_{13} \simeq 1$, we find
\begin{equation}
\sin \theta_{12} \simeq \frac{\sqrt{2} \, \epsilon_{12}}{\sqrt{(\epsilon_{32}-\epsilon_{22})^2 + 2 
\epsilon_{12}^2} },
\label{eq:s12A}
\end{equation}
or equivalently,
\begin{equation}
\tan \theta_{12} \simeq \frac{\sqrt{2} \, \epsilon_{12}}{\epsilon_{32}-\epsilon_{22} }.
\label{eq:t12A}
\end{equation}
Although these relations are similar to those in Eq.~(\ref{eq:s12}) and Eq.~(\ref{eq:t12}) derived in the framework 
of single right-handed neutrino dominance, they are independent on 
any assumption about the dominance of ${\cal M}_1$ or ${\cal M}_2$. 

We see, that we can successfully obtain close to maximal atmospheric mixing angle, large solar mixing angle and 
small 
1-3 mixing angle assuming that both ${\cal M}_1$ and  ${\cal M}_2$ contribute comparably to the left-handed neutrino 
mass matrix. Furthermore, in this case mild hierarchy (or no hierarchy at all) in masses of the heavier two neutrinos 
can be achieved.
We will demonstrate that this scenario works on a simple example. But before we do that let us 
discuss what changes when the contribution from ${\cal M}_3$ is included.

\subsection{Step 3: Mass of the lightest neutrino}

The mass of the lightest neutrino is lifted when ${\cal M}_3$ is taken into
account. Since we assume it is just a small correction to the first two terms it can be treated as a
perturbation. Adding this perturbation does not significantly affect the heavy two eigenvalues and
the diagonalization matrix, but it is crucial for the lightest eigenvalue which is exactly zero in
the limit when this term is ignored.
In the case of non-degenerate eigenvalues, corrections to eigenvalues $m_i$ of a matrix $M$ generated
by a matrix $\delta M$ are given as:
\begin{equation}
\delta m_i = u_i^\dag \, \delta M \, u_i ,
\end{equation}
where $u_i$ are normalized eigenvectors.
In our case:
\begin{equation}
\delta M = - \frac{\lambda^2 v^2_u}{M_0} \, 
\left(\begin{array}{ccc}
   \epsilon_{13}^2 & \epsilon_{13} \epsilon_{23} & \epsilon_{13}  \\
   \epsilon_{23} \epsilon_{13} & \epsilon_{23}^2 & \epsilon_{23}  \\
   \epsilon_{13}  & \epsilon_{23}  & 1
                   \end{array} \right),
\end{equation}
and the vector $\vec v_0$ corresponding to the zero eigenvalue is the first row of $U_{\nu_L}$ obtained from 
${\cal M}_1$ and ${\cal M}_2$. It corresponds to the first column of the lepton mixing matrix $U$, 
see Eq.~(\ref{eq:U}), 
since $U_e \simeq 1$. Therefore, we have
$\vec v_0 \simeq ( U_{e1}, U_{\mu 1},  U_{\tau 1} )^\dag$.
Since $\delta M$ is strongly dominated by the 3-3 element we get the mass of the lightest neutrino in the form 
\begin{equation}
m_{\nu_1} =  \frac{\lambda^2_\nu v^2_u}{M_0} \, | \, U_{\tau 1} \, |^2 ,
\label{eq:m_nu1}
\end{equation}
which does not depend on perturbations ($\epsilon_{ij}$) in the leading order.

Although we do not measure $U_{\tau 1}$, it is related to the observed mixing angles due to the unitarity of the
lepton mixing matrix. In the case $\sin \theta_{13} \simeq 0$ it is simply given as
$U_{\tau 1} \simeq \sin \theta_{23} \, \sin \theta_{12}$.
A global analysis of neutrino oscillation data~\cite{Gonzalez-Garcia:2003}
gives the $3 \sigma$ range:
$0.20 \leq  | \, U_{\tau 1} \, |  \leq 0.58$.

In
simple SO(10)  models $\lambda_u  = \lambda_\nu $, in which case the lightest and the
heaviest fermion of the standard model
are connected through the relation above where $\lambda_\nu^2 v_u^2$ is replaced by
$m_{top}^2$ (actually to be precise, $\lambda_u = \lambda_\nu$ is a relation at the GUT scale and the effects of
the renormalization group running between the GUT scale and the electroweak scale should be taken into account).
This is a very pleasant feature since       
we can further                              
identify $M_0$ with the GUT scale, $M_{GUT} \sim 2 \times 10^{16}$ GeV, in which case we get
\begin{equation}
m_{\nu_1} = \frac{m_{top}^2}{M_{GUT}} \, | \, U_{\tau 1} \, |^2 ,
\label{eq:m_nu1c}
\end{equation}
and predict the mass of the lightest neutrino
to be between $ 5 \times 10^{-5}$ eV and  $ 5 \times 10^{-4}$ eV depending on the value of $U_{\tau 1}$.
This prediction does not depend on details of a model. It represents a realization of Yukawa  coupling
unification in the neutrino sector and adds to predictions of Yukawa coupling unification
in quark and charged lepton sector~\cite{bdr}.

\subsection{A simple example}

Let us demonstrate on a simple example that the scenario discussed in the above sections really works. 
Let us assign the following values to elements of the neutrino Yukawa matrix: $\epsilon_{11} = 0$, 
$\epsilon_{21} = \epsilon_{31} = \epsilon_{12} =\epsilon_{13} = \delta$, $\epsilon_{22} = \epsilon$,
and $\epsilon_{32} = \epsilon_{23} = \epsilon + 2 \delta$, where $\epsilon = 0.01$ and $\delta = 0.002$, 
and let us take $r_1 = 9 \times 10^{-8}$ and $r_2 = -5 \times 10^{-6}$. This choice of parameters 
certainly satisfies the desired 
texture in Eq.~(\ref{eq:texture}). For simplicity we assume a symmetric Yukawa matrix, however it is not 
crucial in any way. For example, exact values of $\epsilon_{23}$, $\epsilon_{13}$ are not relevant at all, and 
would 
not significantly change numerical results below even if changed by a factor of 10. Also $\epsilon_{21}$ does not have to be 
equal 
to $\epsilon_{12}$. The only crucial relations are: $\epsilon_{21} \simeq \epsilon_{31}$ and $\epsilon_{32} \simeq 
\epsilon + 2 \delta$. Finally, let us make the minimal and the most interesting assumption that $M_0 = M_{GUT}$ 
and $\lambda_\nu = \lambda_u$.
With these values of parameters we find:
\begin{eqnarray}
\sin^2 \theta_{23} & = & 0.71 , \\
\sin^2 \theta_{12} & = & 0.35 , \\
\sin^2 \theta_{12} & = & 0.03 ,
\end{eqnarray}
and 
\begin{eqnarray}
m_{\nu_3} & = & 4.9 \times 10^{-2} \; {\rm eV} , \\
m_{\nu_2} & = & 7.8 \times 10^{-3} \; {\rm eV} , \\
m_{\nu_1} & = & 2.5 \times 10^{-4} \; {\rm eV} , \label{eq:m_nu1_num}
\end{eqnarray}
which is in a good agreement with experimental values. The $\sin^2 \theta_{23}$ is somewhat too large (but still 
within $3\sigma$). We are not trying to provide the best fit to data, but rather demonstrate that bi-large mixing 
can be achieved with a simple choice of parameters with the neutrino Yukawa matrix of the form given in 
Eq.~(\ref{eq:texture}).
Note, we parametrized the neutrino Yukawa matrix by two parameters and we 
specified values of all parameters to one digit only. Keeping this in mind, we actually find the results above 
quite remarkable. 
Relaxing the exact relations between elements of the neutrino Yukawa matrix,
there is certainly enough freedom to fit all mixing angles and masses very accurately.
Furthermore, in a realistic scenario one should take into account corrections from the right-handed neutrino mass 
matrix not being exactly diagonal, and also the 
contribution from diagonalization of the charged lepton Yukawa matrix.

Masses of the right-handed neutrinos in this example, $M_3 = M_{GUT}$, $M_2 \simeq 10^{11}$ GeV and $M_1 
\simeq 10^{9}$ GeV, are in an interesting range for leptogenesis~\cite{Pascoli:2003uh}.
The value of $r_2 \, | \, {\vec e}_1 \, |^2 / r_1 \, | \, {\vec e}_2 \, |^2 \simeq -1.5$ measures the 
relative 
contribution of ${\cal M}_1$ and ${\cal M}_2$. It shows that the contributions of the two 
right-handed 
neutrinos 
to the left-handed neutrino mass matrix do not have to be extremely close.

Finally, we can check if  the formula we derived for the mass of the lightest neutrino works. 
Plugging $U_{\tau 1} = 0.41$ 
and $m_{top}^2/M_{GUT} = 0.0014$ eV (the same value used in the evaluation of the masses above) 
to Eq.~(\ref{eq:m_nu1c}) we find $m_{\nu_1} = 2.4 \times 10^{-4}$ eV which is in a very good 
agreement with the numerical result in Eq.~(\ref{eq:m_nu1_num}).

\section{\label{sec:double-see-saw} Origin of strong right-handed neutrino hierarchy}

In SO(10) models the right-handed neutrino is part of  16 dimensional representation. Therefore the Majorana 
mass term is not allowed by SO(10) symmetry and it is generated in the process of  GUT symmetry 
breaking. 
A simple possibility is to assume that a $16$, $\overline{16}$ pair of Higgs fields gets a vev in the 
right-handed neutrino direction. In models where the two Higgs doublets originate from 10 dimensional representation 
of SO(10), Yukawa couplings are generated from operators of the form
\begin{equation}
16_i \, (...)_{ij} \, 10 \, 16_{j}, \quad \quad i= 1,2,3, 
\end{equation}
where $10$ contains the two Higgs doublets of MSSM, and $(...)_{ij}$ 
is a flavor dependent part responsible for generating the desired structure of 
Yukawa matrices. It typically contains flavon fields responsible for generating hierarchy between generations 
and 
other Higgs fields, for example 45 of SO(10), responsible for the right quark-lepton mass relations of the first two 
generations. 
Besides these, there can also be operators where $10$ of Higgs and one $16_i$ are 
replaced by $\langle \overline{16} \rangle_{\bar \nu}$ and  SO(10) singlet fields $N_i$ of the form 
\begin{equation}
16_i \, (...)_{ij} \, {\langle \overline{16} \rangle}_{\bar \nu} \, N_{j}. 
\end{equation}
These operators generate a mass matrix, $Y_N$,  
between 
right-handed neutrinos and SO(10) singlets. It has naturally the same  structure as other Yukawa matrices, although it is 
not identical, because $(...)_{ij}$ distinguishes between different fields in $16$ dimensional representation. 
A mass term for singlet fields effectively leads to a 
Majorana mass matrix for right-handed neutrinos:
\begin{equation}
M_{\nu_R} = Y_N \ M_N^{-1} \ Y_N^T .    
\end{equation}
A simple mass term for singlet fields of the form,
$M_N \simeq  {\rm diag} (1,1,1)M_N$,
where for simplicity we use $M_N$ for both the matrix and the scale itself, automatically leads to a strong hierarchy in 
masses of right-handed neutrinos. If the hierarchy in mass eigenvalues of $Y_N$ is $\sim
(\delta_N, \epsilon_N, 1)$ (similar to eigenvalues of other Yukawa matrices) the hierarchy in right-handed neutrino masses 
is  naturally doubled 
$\sim (\delta_N^2, \epsilon_N^2, 1)$.

The left-handed neutrino Majorana mass term can be written as:
\begin{equation}
M_{\nu_L} = Y_{\nu} \ Y_N^{T \, -1} \ M_N \ Y_N^{-1} \ Y_{\nu}^T .
\end{equation}
If $Y_N$ is identical to $Y_\nu$ then $M_{\nu_L} = M_N$ and all three neutrinos have almost the same mass.
This is to demonstrate that it is actually very natural to expect that all three right-handed neutrinos 
contribute roughly equally to the resulting left-handed neutrino mass matrix. The desired situation when 
$M_1$ and $M_2$ contribute more is possible to achieve in two ways: either to manage $\epsilon_N$ 
and $\delta_N$ to be somewhat smaller than the corresponding perturbations in $Y_\nu$,
or simply assume 
that $M_{N_1} \simeq M_{N_2} > M_{N_3}$. From model building point of view, the second choice  
does not represent a big challenge. As a by-product, the stronger hierarchy of $M_{\nu_R}$ makes 
it
also closer to the identity matrix, and  corrections to the discussion in previous sections from it not 
being exactly diagonal are smaller.

\section{\label{sec:conclusions} Conclusions}

We discussed conditions under which bi-large lepton mixing can be achieved in hierarchical models in which all mass
matrices are dominated by the 3-3 element.
Many features of this framework are similar to those in the democratic framework discussed in Ref.~\cite{dem1}.
The obvious quark-lepton symmetry makes it easy to embed models of this type into GUTs.
The right-handed neutrino mass scale can
be identified with the GUT scale in which case
the mass of the lightest neutrino
is given as $(m_{top}^2/M_{GUT}) \, | \, U_{\tau 1} \, |^2$, the same as in Ref.~\cite{dem1}.
The third generation Yukawa coupling unification is obvious in this picture, since it is our starting point and we do not 
allow any large off-diagonal elements. This can be understood from two possible ways  permutation symmetry can be used in
model building. A matrix with 3-3 element only can be also motivated by permutation symmetry  under
which the first two families transform as a doublet~\cite{D3}.
Both approaches require strong hierarchy in masses of right handed neutrinos and negligible contribution of the heaviest one to 
the left-handed neutrino mass matrix.

There are few major differences however. There is a tension between achieving the observed spectrum of heavier two 
neutrinos and 
bi-large mixing in the hierarchical approach, which is avoided when the two lighter right-handed neutrinos contribute comparably 
to the 
left-handed neutrino mass matrix. In the democratic approach there is no tension at all, since bi-large mixing originates 
predominantly from the matrix diagonalizing the charged lepton Yukawa matrix and so neutrino masses can be adjusted arbitrarily.
Furthermore, in the democratic approach, there is a well defined
framework (without exactly specifying perturbations) in which the left-handed neutrino mass matrix contributes the
minimal amount of mixing to the lepton mixing
matrix and the value of
one mixing angle, $\sin \theta_{13}$, can be predicted~\cite{dem1}. 
We do not see the equivalent situation in the hierarchical approach.
Finally, it seems to be much easier to build  concrete GUT models with hierarchical Yukawa matrices of the type discussed here 
than democratic ones.

\vspace{0.5cm}
\noindent
{\bf Acknowledgments:} 
I would like to thank S. Pascoli, S. Raby and  S. Nussinov
for useful comments and discussions.
This work was supported, in part, by the U.S.\ Department of Energy, Contract
DE-FG03-91ER-40674 and the Davis Institute for High Energy Physics.


\end{document}